\documentstyle[12pt]{article}

\catcode`@=12
\begin{document}
\def\tilde{\widetilde}
%\def\thefootnote{\fnsymbol{footnote}}
%\FERMILABPub{94/119--T}
\begin{titlepage}
\begin{flushright}
        FERMILAB--PUB--94/119--T\\
        OSU Preprint 289\\
        May 1994\\
\end{flushright}
%\vspace{-0.20in}
\begin{center}
{\large \bf Comparison of SO(10)-Symmetric Fermion Mass Matrices\\
	with and without Degenerate Neutrinos}\\
%\vfill
\vskip 0.20in
        {\bf Carl H. ALBRIGHT}\\
 Department of Physics, Northern Illinois University, DeKalb, Illinois
60115\footnote{Permanent address}\\[-0.2cm]
        and\\[-0.2cm]
 Fermi National Accelerator Laboratory, P.O. Box 500, Batavia, Illinois
60510\footnote{Electronic address: ALBRIGHT@FNALV}\\
        and\\
        {\bf Satyanarayan NANDI}\\
 Department of Physics, Oklahoma State University, Stillwater, Oklahoma
        74078\footnote{Electronic address: PHYSSNA@OSUCC}\\
\end{center}
%\vskip 0.5in
\vfill
\begin{abstract}
It has been recently suggested by others that one can simultaneously
explain the depletions of solar electron-neutrinos and atmospheric
muon-neutrinos along with a 7~eV neutrino component of mixed dark matter
by postulating the existence of nearly-degenerate 2 eV neutrinos with
the correct mixing parameters.  We study this claim in the framework of
a simple SO(10)-symmetric model constructed from the low scale data using
a bottom-up procedure recently advanced by the authors and compare the
results with and without degenerate neutrinos.
\end{abstract}
\end{titlepage}

In the past several years, three sets of observations, two direct and one
indirect, suggest that neutrinos have mass with their masses and mixing
parameters lying in restricted regions of the mixing planes according to
certain user-preferred interpretations.
\begin{itemize}
\item[1)] The depletion of solar neutrinos has now been observed\cite{solar}
	in experiments
	sensitive to the p-p, $^7Be$ and $^8B$ electron-neutrinos with the
	preferred\cite{solarint} particle physics interpretation that these
	neutrinos undergo
	resonant conversion into muon-neutrinos in passing through the dense
	solar matter.  This non-adiabatic Mikheyev-Smirnov-Wolfenstein (MSW)
	effect\cite{MSW} is restricted to the
	central region of the 12 mixing plane where
	$$ \delta m^2_{12} = 5 \times 10^{-6}\ {\rm eV}^2, \qquad
		\sin^2 2\theta_{12} = 8 \times 10^{-3}\eqno(1a)$$
\item[2)] The apparent depletion of the atmospheric muon-neutrino flux relative
	to the atmospheric electron-neutrino flux has now been
	observed\cite{atm} by three experimental collaborations and is
	widely interpreted as due to the oscillation of
	muon-neutrinos into tau-neutrinos during their passage through the
	atmosphere.\cite{solarint}  Although this phenomenon is considered to
	be on more shaky ground than the solar depletion effect, several recent
	refined flux calculations\cite{GP} have further restricted the
	mixing region but have not ruled out this interpretation.  A typical
	point in the 23 mixing plane is represented by
	$$\delta m^2_{23} = 2 \times 10^{-2}\ {\rm eV}^2, \qquad
		\sin^2 2\theta_{23} = 0.5 \eqno(1b)$$
\item[3)] The indirect evidence for neutrino mass is strengthened somewhat
	by the recent COBE observation\cite{COBE} of density fluctuations in
	the universe, for which the cocktail model\cite{cocktail} attributes
	the 30\% hot dark matter component to neutrinos provided
	$$ \sum_i m_{\nu,i} \simeq 7\ {\rm eV} \eqno(1c)$$
	If the tau-neutrino is naturally assumed to be the heaviest with
	a mass of 7 eV, the present accelerator data\cite{accdata} limit the
	mixing in the 23 plane to the region
	$$ m_{\nu_{\tau}} \simeq 7\ {\rm eV}, \qquad
		\sin^2 2\theta_{23}\ \ltap\ 4 \times 10^{-3}\eqno(1d)$$
\end{itemize}

It is clear from the above that one can not explain all three effects with the
interpretations cited in the framework of three non-degenerate, light
left-handed neutrinos.  One possibility is to introduce a new light
sterile neutrino,\cite{sterile} but a more economical recent
suggestion\cite{degsugg} makes the assumption
the three light neutrinos are nearly degenerate with masses close to 2 eV.
In this way, the three neutrinos mix to give the solar and atmospheric
depletions with the correct $\delta m_{ij}^2$'s, while they share equally
in providing the hot component of mixed dark matter.

In this short note, we extend our recent construction\cite{AN} of
SO(10)-symmetric
mass matrix models for quarks and leptons to the case of degenerate
neutrinos and compare our results with the non-degenerate cases considered
earlier.  Our bottom-up procedure allows us to start from the complete set
of quark and lepton masses and two mixing matrices defined at the low scales
and to reconstruct numerically quark and lepton mass matrices at the
grand unification scale which yield the low scale results.  By choosing the
bases judiciously, we can single out mass matrices which exhibit simple SO(10)
structure with as many texture zeros as possible from which simple mass
matrix models can be identified.  It is interesting to note that our findings
in the non-degenerate case pointed to a greater simplicity for the mass matrix
model incorporating observations 1) and 2) above rather than 1) and 3).

We shall summarize briefly the input and
procedure and refer the interested reader to Ref. \cite{AN} for more of the
details. From the information given in (1a) and (1b), we
have taken for the lepton input in the non-degenerate (ND) case
$$\begin{array}{rlrl}
        m^{ND}_{\nu_e}&= 0.5 \times 10^{-6}\ {\rm eV},& \qquad m_e&= 0.511\
		{\rm MeV}\nonumber\\
        m^{ND}_{\nu_{\mu}}&= 0.224 \times 10^{-2}\ {\rm eV},& \qquad m_{\mu}&=
                105.3\ {\rm MeV}\cr
        m^{ND}_{\nu_{\tau}}&= 0.141\ {\rm eV},& \qquad m_{\tau}&= 1.777\ {\rm
                GeV}\cr \end{array}\eqno(2a)$$
and
$$V_{LEPT} = \left(\matrix{0.9990 & 0.0447 & (-0.690 -0.310i)
		\times 10^{-2}\cr -0.0381 -0.0010i & 0.9233 & 0.3821\cr
                0.0223 -0.0030i & -0.3814 & 0.9241\cr}\right) \eqno(2b)$$
where we have assumed a value of for the electron-neutrino mass to
which our analysis is not very sensitive and constructed
the lepton mixing matrix by making use of the unitarity conditions.
In the degenerate (DEG) case, we shall leave the mixing matrix unchanged and
simply replace the set of light neutrino masses above by the new set
$$\begin{array}{rlrl}
        m^{DEG}_{\nu_e}&= (2.0 - 0.125 \times 10^{-5})\ {\rm eV}\nonumber\\
        m^{DEG}_{\nu_{\mu}}&= 2.0\ {\rm eV}\cr
        m^{DEG}_{\nu_{\tau}}&= (2.0 + 0.500 \times 10^{-2})\ {\rm eV}\cr
		\end{array}\eqno(2c)$$
which yield the correct $\delta m^2_{ij}$'s with $\sum_i m_{\nu,i} \simeq 6$
eV.

For the quark input data, we evaluated the light quark masses at 1 GeV and the
heavy quark masses at their running masses according to
$$\begin{array}{rlrl}
        m_u(1 {\rm GeV})&= 5.1\ {\rm MeV},& \qquad m_d(1 {\rm GeV})&= 8.9\
                {\rm MeV}\nonumber\\
        m_c(m_c)&= 1.27\ {\rm GeV},& \qquad m_s(1 {\rm GeV})&= 175\ {\rm MeV}
                \cr
        m_t(m_t)&= 150\ {\rm GeV},& \qquad m_b(m_b)&\simeq 4.25\ {\rm GeV}\cr
  \end{array}\eqno(3a)$$
corresponding to a pole mass of $m_t^{phys} \sim 160$ GeV for the top quark.
We adopted the following central values for the Cabibbo-Kobayashi-Maskawa
(CKM) mixing matrix\cite{PDB} at the weak scale
$$V_{CKM} = \left(\matrix{0.9753 & 0.2210 & (-0.283 -0.126i)\times 10^{-2}\cr
                -0.2206 & 0.9744 & 0.0430\cr
                0.0112 -0.0012i & -0.0412 -0.0003i & 0.9991\cr}\right)
        \eqno(3b)$$
where we assumed a value of 0.043 for $V_{cb}$ and applied strict
unitarity to determine $V_{ub},\ V_{td}$ and $V_{ts}$.

The masses and mixing matrices were then evolved to the GUT scale by means
of the one-loop renormalization group equations\cite{Nac} for the minimal
supersymmetric
standard model (MSSM).  By means of Kusenko's method\cite{Kus} extended to
leptons as
well as quarks, one can then construct complex symmetric quark, charged lepton
and light neutrino mass matrices which follow from the information at the GUT
scale.  The variation of two parameters controlling the choice of quark and
lepton bases enabled us to identify mass matrices which exhibit simple
SO(10) symmetry with a maximum number of texture zeros.  The up, down, charged
lepton and Dirac neutrino mass matrices in the
non-degenerate case received contributions from two ${\bf 10}$'s and two
${\bf 126}$'s of SO(10) according to
$$\begin{array}{rl}
	M^U&\sim M^{N_{Dirac}} \sim diag(126;\ 126;\ 10)\nonumber\\[0.1in]
	M^D&\sim M^E \sim \left(\matrix{10',126 & 10',126' & 10'\cr 10',126'
		& 126 & 10'\cr  10' & 10' & 10\cr}\right)\cr
	\end{array}\eqno(4)$$
{}From the numerical results and these simple forms, the following matrix
model emerged
$$\begin{array}{rlrl}
        M^U&= diag(F',\ E',\ C')& \qquad M^{N_{Dirac}}&= diag(-3F',\ -3E',\ C')
		\nonumber\\[0.1in]
        M^D&= \left(\matrix{0 & A & D\cr A & E & B\cr D & B & C\cr}\right)
		& \qquad
        M^E&= \left(\matrix{F & 0 & D\cr 0 & -3E & B\cr D & B & C\cr}\right)
		\end{array}\eqno(5)$$
where only $D$ is complex and the constraint, $4F'/F = -E'/E$, holds.  There
are nine independent parameters present along with four texture zeros for both
the quark and lepton mass matrices.

If we now repeat the calculation for the degenerate case, replacing the input
in (2a) by that in (2c), we find exactly the same numerical results for the
up, down and charged lepton mass matrices with the same choice of bases; only
the light or effective neutrino mass matrix is altered.  The results for the
non-degenerate and degenerate cases are, respectively,
$$\begin{array}{rl}
M^{N_{eff}}_{ND} = \left(\matrix{(.4839 + .1534i)\times 10^{-4}&
	(-.9059 - .1304i)\times 10^{-3}& (.3023 + .0374i)\times 10^{-2}\cr
	(-.9059 - .1304i)\times 10^{-3}& (.1465 - .0001i)\times 10^{-1}&
	(-.5065 + .0002i)\times 10^{-1}\cr
	(.3023 + .0374i)\times 10^{-2}& (-.5065 + .0002i)\times 10^{-1}&
	0.1502\cr}\right)\cr\end{array}\eqno(6a)$$
and
$$\begin{array}{rl}
M^{N_{eff}}_{DEG} = \left(\matrix{2.322 &
	(-.4714 + .0152i)\times 10^{-2}& (.0180 + .1202i)\times 10^{-1}\cr
	(-.4714 + .0152i)\times 10^{-2}& 2.370 &
	(-.1832 + .0225i)\times 10^{-2}\cr
	(.0180 + .1202i)\times 10^{-1}& (-.1832 + .0225i)\times 10^{-2}&
	2.3752\cr}\right)\cr\end{array}\eqno(6b)$$
in units of electron volts.

A striking difference is observed between the two matrices.  While a hierarchy
similar to that for the down quark or charged lepton mass matrix is found in
the non-degenerate case, when the neutrinos are nearly degenerate a larger
diagonal matrix proportional to the identity is superposed on a smaller and
more uniform background matrix.  The conventional interpretation in the
non-degenerate case is the operation of a seesaw mechanism,\cite{GMRS}
whereby an extremely large right-handed Majorana neutrino mass matrix $M^R$
gives rise to the observed light neutrinos according to
$$M^{N_{eff}}_{ND} = - M^{N_{Dirac}}(M^R)^{-1}M^{N_{Dirac}T}\eqno(7a)$$
The suggestion by a number of authors\cite{degsugg} in the degenerate case is
that, an additional diagonal and family-independent contribution arises from a
left-handed Majorana neutrino mass matrix according to
$$M^{N_{eff}}_{DEG} = M^L - M^{N_{Dirac}}(M^R)^{-1}M^{N_{Dirac}T}\eqno(7b)$$
Although such a term has been disfavored since it requires a Higgs triplet
at tree level, the new suggestion is that $M^L$ arises through a
nonrenormalizable dimension-five operator of the form
	$$M^L = m_LI = \lambda_L\nu_L^T{<\phi>^2\over{M}}\nu_L \eqno(8)$$
which yields the right order of magnitude with
$M$ one of the heavy right-handed Majorana masses
and $<\phi>$ the Higgs expectation value at the weak symmetry-breaking scale.

If we now apply (7a) or (7b) making use of (6a) or (6b),
respectively, and the diagonal Dirac neutrino matrix suggested by the model
matrices in (4); we can determine the required right-handed Majorana matrix
$M^R$ for each scenario.  From Ref. \cite{AN} where the best values for
$C',\ E'$ and $F'$ are given, we find numerically
$$ M^R_{ND} = \left(\matrix{(.1744 - .0044i)\times 10^{10} &
	(-.2332 + .0153i)\times 10^{11} & (-.2811 - .1925i)\times 10^{12}\cr
	(-.2332 + .0153i)\times 10^{11} & (.6773 - .0329i)\times 10^{12} &
	(-.1189 + .0243i)\times 10^{14}\cr
	(-.2811 - .1925i)\times 10^{12} & (-.1189 + .0243i)\times 10^{14} &
	(.6045 + .0624i)\times 10^{15}\cr}\right) \eqno(9a)$$
for the non-degenerate case and
$$ M^R_{DEG} = \left(\matrix{(-.4320 - .1444i)\times 10^{7} &
	(.1227 + .0253i)\times 10^{9} & (-.0730 + .3750i)\times 10^{11}\cr
	(.1227 + .0253i)\times 10^{9} & (-.2185 - .0032i)\times 10^{11} &
	(.0158 - .1057i)\times 10^{13}\cr
	(-.0730 + .3750i)\times 10^{11} & (.0158 - .1057i)\times 10^{13} &
	(.4634 + .1903i)\times 10^{14}\cr}\right) \eqno(9b)$$
for the degenerate case in units of GeV.  We have used a value of $m_L = 2.322$
eV at the GUT scale in order to minimize the hierarchy in the eigenvalues of
$M^R_{DEG}$.  The eigenvalues of the above matrices for the two cases are
then found to be
$$\begin{array}{rlrl}
M^{R_1}_{ND}&= 0.213\times 10^9\ {\rm GeV} &\qquad\qquad
	M^{R_1}_{DEG}&= 0.165 \times 10^9\ {\rm GeV} \nonumber\\
M^{R_2}_{ND}&= 0.475\times 10^{12}\ {\rm GeV} &\qquad\qquad
	M^{R_2}_{DEG}&= 0.277\times 10^{10}\ {\rm GeV} \cr
M^{R_3}_{ND}&= 0.608\times 10^{15}\ {\rm GeV} &\qquad\qquad
	M^{R_3}_{DEG}&= 0.501\times 10^{14}\ {\rm GeV} \cr
	\end{array}\eqno(10a)$$
If one uses (4), (6b) and (7a) without the diagonal $M_L$ contribution,
a much larger hierarchy appears in the degenerate case given by
$$\begin{array}{rl}
M^{R_1}&= 0.469 \times 10^4\ {\rm GeV} \nonumber\\
M^{R_2}&= 0.469\times 10^{9}\ {\rm GeV} \cr
M^{R_3}&= 0.490\times 10^{13}\ {\rm GeV} \cr \end{array}\eqno(10b)$$

Another interesting observation pertains to the structures of the matrices
in (9a) and (9b).  The matrix for the non-degenerate scenario was observed
in Ref. \cite{AN} to have a near geometric texture,\cite{Lemke} i.e.,
a superposition
of two geometric forms (with some elements zero), which suggests two ${\bf
126}$
contributions, as approximated by
$$M^R_{ND} = \left(\matrix{F'' & - {2\over{3}}\sqrt{F''E''} &
                -{1\over{3}}\sqrt{F''C''}e^{i\phi_{D''}}\cr
                - {2\over{3}}\sqrt{F''E''} & E'' &
                        -{2\over{3}}\sqrt{E''C''}e^{i\phi_{B''}}\cr
                -{1\over{3}}\sqrt{F''C''}e^{i\phi_{D''}} &
                -{2\over{3}}\sqrt{E''C''}e^{i\phi_{B''}} & C''\cr}\right)
                        \eqno(11a)$$
where $E'' = {2\over{3}}\sqrt{F''C''}$ and $\phi_{B''} = - \phi_{D''}/3$.
For the degenerate case with $m_L = 2.322$ eV on the other hand, we find after
a rephasing of the matrix
$$M^R_{DEG} = \left(\matrix{F'' & - 0.4\sqrt{F''E''} &
                - 2.5\sqrt{F''C''}e^{i\phi_{D''}}\cr
                - 0.4\sqrt{F''E''} & E'' & \sqrt{E''C''}e^{i\phi_{B''}}\cr
                - 2.5\sqrt{F''C''}e^{i\phi_{D''}} &
                \sqrt{E''C''}e^{i\phi_{B''}} & - C''\cr}\right)
                        \eqno(11b)$$
where $E'' = 1.4\sqrt{F''C''}$, with $\phi_{D''}$ and $\phi_{B''}$ near
$\pm 90^o$.  However, the simple superposition structure
of two geometric forms is lost.  This situation can only be improved somewhat
by varying $m_L$ at the expense of introducing a much larger hierarchy in
the eigenvalues for $M^R_{DEG}$.  In the
non-degenerate case, only three additional parameters need be added to the
nine already counted in the model, while in the degenerate case,
seven parameters ($m_L$, four magnitudes and two phases) must be added.

Several other authors have recently analysed\cite{deganal} the
degenerate case, but they employed real diagonal matrices for the up
quark, Dirac neutrino and right-handed Majorana neutrino mass matrices
resulting in no simple SO(10) Higgs structure at the GUT scale.
For their choice of matrices, eighteen parameters are required if the proper
phases are included.
In this note we have demonstrated that a mass matrix model can be constructed
with fewer parameters,
not only for the non-degenerate neutrino scenario but for the degenerate one
as well, which exhibits a simple SO(10) structure with four texture zeros
in the quark and in the lepton mass matrices.  The simplicity arises in the
basis where the up quark, and hence Dirac neutrino, mass matrices are
diagonal, but the light neutrino and right-handed Majorana matrices are
non-diagonal and complex-symmetric.\cite{KS}

The basic change in going from the non-degenerate to degenerate case arises
in the light neutrino mass matrix at the GUT scale, which suggests the
presence of a family-independent, family-diagonal left-handed Majorana
matrix and a modified heavy right-handed Majorana mass matrix.
But the observed superposition of two ${\bf 126}$ geometric matrices
for the right-handed Majorana matrix is lost with the introduction of at
least four more parameters required than in the non-degenerate case.

These complicating features suggest to us that the non-degenerate scenario
where the solar and atmospheric neutrino depletions are readily understood
in the simplest SO(10) framework with the fewest number of parameters remains
the most viable theoretical one at present, provided the COBE
results\cite{COBE} can
be interpreted entirely in terms of cold dark matter.  Future experiments which
can further explore the limits on double beta decay\cite{betabeta} and can
heroically lower the present upper limit of 7.3 eV on the electron-neutrino
mass\cite{emass}
by one order of magnitude will definitively clarify this issue.

The research of CHA was supported in part by Grant No. PHY-9207696 from the
National Science Foundation, while that of SN was supported in part by the
U.S. Department of Energy, Grant No. DE-FG05-85ER 40215.\\

\end{document}